\documentstyle[12pt]{article}
\newcommand{\be}{\begin{equation}}
\newcommand{\ee}{\end{equation}}
\newcommand{\bea}{\begin{eqnarray}}
\newcommand{\eea}{\end{eqnarray}}
\newcommand{\norsl}{\normalsize\sl}
\newcommand{\norsc}{\normalsize\sc}
\begin{document}

\begin{titlepage}

\title{ Spin Structure Function $g_2(x,Q^2)$ and \\
           Twist-3 Operators in large-$N_C$ QCD
               }
\author{
\norsc  Ken SASAKI\thanks{e-mail address: sasaki@ed.ynu.ac.jp}\\
\norsl  Dept. of Physics,  Faculty of Engineering, Yokohama National University\\
\norsl  Yokohama 240-8501, JAPAN\\}

\date{}
\maketitle
 
\begin{abstract}
{\normalsize It is shown in the framework of the operator product expansion and the 
renormalization group method that the twist-3 part of flavour nonsinglet spin 
structure function $g_2(x,Q^2)$ obeys a simple 
Dokshitzer-Gribov-Lipatov-Altarelli-Parisi (DGLAP) equation in the large $N_C$ limit 
even in the case of massive quarks ($N_C$ is the number of colours). 
There are four different types of twist-3 operators which contribute to $g_2$, 
including  quark-mass-dependent operators and the ones proportional 
to the equation of motion. They are not all 
independent but are constrained by one relation. A new choice of the independent 
operator bases leads to a simple form of the evolution equation for 
$g_2$ at large $N_C$.
}
\end{abstract}
 
\begin{picture}(5,2)(-290,-500)
\put(2.3,-125){YNU-HEPTh-98-101}
\put(2.3,-140){March 1998}
\end{picture}

\thispagestyle{empty}
\end{titlepage}
\setcounter{page}{1}
\baselineskip 18pt

In the experiments of the polarized deep inelastic leptoproduction 
we can obtain the information on spin structures of nucleon, which 
are described by the two functions $g_1(x,Q^2)$ and 
$g_2(x,Q^2)$. The QCD effects on $g_1$ and $g_2$ have been extensively 
studied~\cite{review} since earlier papers~\cite{ARS}-\cite{KOD2}. 
Increasingly accurate  measurements of $g_1$ have been performed at SLAC, CERN 
and DESY~\cite{g1}, while the $g_2$ measurements still have limited statistical 
precision~\cite{g2}.

In the language of the operator-product-expansion (OPE) 
the twist-2 operators contribute to $g_1$ in the leading order of $1/Q^2$. 
As for the structure function $g_2$, on the other hand, both  
twist-2 and twist-3 operators participate   
in the leading order. Moreover, the number of participating twist-3 operators 
grows with spin (moment of $g_2$).  
Due to increase of the number of operators and the 
mixing among these operators the analysis of the twist-3 part of 
$g_2$ turns out to be rather complicated~\cite{ShuVain}-\cite{KMSU}. 
In other words, the $Q^2$ evolution equation 
for the moments of the twist-3 part of $g_2$ cannot be written in a simple 
form but in a sum of terms, the number of which increases with spin. 

For the case of the twist-3 flavour nonsinglet $g_2$
it has been observed by Ali, Braun and Hiller (ABH)~\cite{ABH} that 
in the large $N_C$ limit $g_2$ obeys a simple 
Dokshitzer-Gribov-Lipatov-Altarelli-Parisi (DGLAP) equation~\cite{GLAP}. 
In their formalism of working directly with the nonlocal operator 
contributing to the twist-3 part of $g_2$, they showed that local operators 
involving gluons effectively decouple from evolution equation for large $N_C$.
In fact their analysis has been made with massless quarks. 

In this paper I reanalyze the $Q^2$ evolution of the flavour nonsinglet 
twist-3 part of $g_2$ in the framework of the standard OPE and 
the renormalization group (RG) with massive quarks. 
Actually the OPE analysis of $g_2$ has been performed already and 
the anomalous dimensions of the relevant twist-3 operators have been 
calculated~\cite{BKL}\cite{Ratcliffe}\cite{JiChou}\cite{KUY}\cite{KTUY}. 
However, to the best of my knowledge, the large $N_C$ limit of $g_2$ has not been 
thoroughly studied so far in OPE and RG.
There are four different types of twist-3 operators which contribute to $g_2$, 
including  quark-mass-dependent operators and the ones proportional 
to the equation of motion. They are not all 
independent but are constrained by one relation. 
It was pointed out recently by Kodaira, Uematsu and Yasui~\cite{KUY} that any choice 
of the independent operator bases leads to a unique prediction for the moments. 
Taking a new basis of the independent operators, I will show that 
the $Q^2$ evolution of the twist-3 part of $g_2$ obeys a simple DGLAP equation 
in the $N_C \rightarrow \infty$ limit and thus the ABH result on $g_2$ 
is reproduced even with massive quarks.

The spin structure function 
$g_2$ receives contributions from both twist-2 and twist-3 operators. 
However, the twist-2 part of $g_2$ can be extracted 
once $g_1$ is measured~\cite{WW}:
\be
   g^{tw.2}_2(x,Q^2)=-g_1(x,Q^2) + \int^{1}_{x}\frac{g_1(y,Q^2)}{y}dy  ~.
\ee
Thus the difference
\be
    \overline{g}_2(x,Q^2)=g_2(x,Q^2)-g^{tw.2}_2(x,Q^2)
\ee
contains the twist-3 contributions only.

The twist-3 operators which enter the OPE for 
the flavour nonsinglet $\overline{g}_2$ are the following 
(I follow the notation and 
conventions of Refs.\cite{KUY}\cite{KTUY} and omit the flavour matrices $\lambda_i$):
\bea
  R_F^{\sigma\mu_{1}\cdots \mu_{n-1}} &=&
         \frac{i^{n-1}}{n} \Bigl[ (n-1) \overline{\psi}\gamma_5
       \gamma^{\sigma}D^{\{\mu_1} \cdots D^{\mu_{n-1}\}}\psi
\nonumber\\
   & & \quad - \sum_{l=1}^{n-1} \overline{\psi} \gamma_5
       \gamma^{\mu_l }D^{\{\sigma} D^{\mu_1} \cdots D^{\mu_{l-1}}
            D^{\mu_{l+1}} \cdots D^{\mu_{n-1}\}}
                             \psi \Bigr] -(\rm{traces}) , \label{quark}\\
  R_l^{\sigma\mu_{1}\cdots \mu_{n-1}} &=& \frac{1}{2n}
              \left( V_l - V_{n-1-l} + U_l + U_{n-1-l} \right) , 
      \qquad (l=1,\cdots,n-2)       \label{gluon}\\
  R_m^{\sigma\mu_{1}\cdots \mu_{n-1}} &=&
          i^{n-2} m S' \overline{\psi}\gamma_5
       \gamma^{\sigma}D^{\mu_1} \cdots D^{\mu_{n-2}}
        \gamma ^{\mu_{n-1}} \psi -(\rm{traces}), \label{mass} \\
 R_{E}^{\sigma\mu_{1}\cdots \mu_{n-1}} &=&
           i^{n-2} \frac{n-1}{2n} S' [ \overline{\psi} \gamma_5
          \gamma^{\sigma} D^{\mu_1} \cdots D^{\mu_{n-2}}
        \gamma ^{\mu_{n-1}} (i\not{\!\!D} - m )\psi \nonumber\\
    & & \qquad\qquad \quad + \overline{\psi} (i\not{\!\!D} - m )
              \gamma_5 \gamma^{\sigma} D^{\mu_1} \cdots D^{\mu_{n-2}}
        \gamma ^{\mu_{n-1}} \psi ] -(\rm{traces}),  \label{motion}
\eea
where $\{\ \ \  \}$ means complete symmetrization over the Lorentz indices and 
$m$ represents the quark mass.  The symbol $S'$ denotes symmetrization on the 
indices $\mu_1\mu_2 \cdots \mu_{n-1}$ and antisymmetrization on $\sigma\mu_i$. 
The operators in Eq.(\ref{gluon}) contain
the gluon field strength $G_{\mu\nu}$ and its dual tensor
$\widetilde{G}_{\mu \nu}={1\over
2}\varepsilon_{\mu\nu\alpha\beta}
G^{\alpha\beta}$ and they are given by
\bea
    V_l &=&-i^n g S' \overline{\psi}\gamma_5
       D^{\mu_1} \cdots G^{\sigma \mu_l } \cdots D^{\mu_{n-2}}
        \gamma ^{\mu_{n-1}} \psi -(\rm{traces}) , \label{VV}\\
    U_l &=& i^{n-1} g S' \overline{\psi}
       D^{\mu_1} \cdots \widetilde{G}^{\sigma \mu_l } \cdots
             D^{\mu_{n-2}} \gamma ^{\mu_{n-1}} \psi -(\rm{traces}),  \label{UU}
\eea
where $g$ is the QCD coupling constant. The operator $R_E^n$ in Eq.(\ref{motion}) 
is proportional to the equation of motion (EOM operator). 
The above twist-3 operators are not 
all independent but they are constrained by the following 
relation~\cite{ShuVain}\cite{Jaffe}:
\be
     R_F^{\sigma\mu_{1}\cdots \mu_{n-1}} =
        \frac{n-1}{n} R_m^{\sigma\mu_{1}\cdots \mu_{n-1}}
             + \sum_{l=1}^{n-2} (n-1-l)
                 R_l^{\sigma\mu_{1}\cdots \mu_{n-1}} +
             R_{E}^{\sigma\mu_{1}\cdots \mu_{n-1}} .
\label{oprelation}
\ee
Thus in total there are $n$ independent operators contributing to 
the ($n$-1)-th moment of $\overline{g}_2$. 
But we will see later that in the $N_C \rightarrow \infty$ limit the ($n$-1)-th moment 
is expressed in terms of one operator $R_F^{\sigma\mu_{1}\cdots \mu_{n-1}}$.

In all the analyses of $\overline{g}_2$ performed so far in the framework of OPE and RG, 
operators $R_l$,$R_m$,$R_E$ of Eqs.(\ref{gluon})-(\ref{motion}) have been taken as 
independent bases.  In this paper I choose $R_F$,$R_l$,$R_E$ as independent 
operators, replacing $R_m$ with $R_F$ of Eq.(\ref{quark}). 
The advantage of this choice of operator basis is that the coefficient functions 
take simple forms at the tree-level.  In fact we have~\cite{KUY} 
\be
   E^n_F({\rm tree})=1, \qquad  E^n_l({\rm tree})=0 \quad {\rm for}\  l=1,\cdots,n-2
\ee
since the anti-symmetric part of the short distance expansion 
for the product of two electromagnetic currents can be written at the tree level as 
\bea
 & & i\int d^4xe^{iq\cdot x} T(J_{\mu}(x)J_{\nu}(0))\vert_{\rm anti-symmetric} \nonumber\\
 & & \qquad = - i\varepsilon_{\mu\nu\lambda\sigma} q^{\lambda}
         \sum_{n=1,3,\cdots} \left( \frac{2}{Q^2 }\right )^n
            q_{\mu _1} \cdots q_{\mu _{n-1}}
       \bigl \{ R_q^{\sigma\mu_1\cdots\mu_{n-1}} + R_F^{\sigma\mu_1
                          \cdots\mu_{n-1}} \bigr \}   \nonumber  \\
 & & \qquad \quad \cdots , \label{extope}
\eea
where dots $\cdots$ stands for non-leading terms and 
\noindent
\be
   R_q^{\sigma\mu_1\cdots\mu_{n-1}}=i^{n-1} \overline{\psi}\gamma_5
    \gamma^{\{{\sigma}}D^{\mu_1} \cdots D^{\mu_{n-1}\}}\psi
          -(\rm{traces})
\ee
are  twist-2 operators which contribute to $g_1$ and $g_2^{tw.2}$. 
It is true that due to the relation, Eq.(\ref{oprelation}), 
$R_F^{\sigma\mu_1\cdots\mu_{n-1}}$ 
can be expressed in terms of other operators. When eliminating $R_F^n$, we obtain 
a different set of coefficient functions.  In other words, 
the (tree-level) coefficient functions are dependent upon the choice 
of the independent operators~\cite{KUY}.

The renormalization constants for this new set of independent operators 
are written in the matrix form as
\bigskip
\bea
\left(\matrix{R^n_F\cr
              R^n_l\cr
              R^n_E\cr}\right)_B =
\left(\matrix{{\widetilde Z}_{FF}&{\widetilde Z}_{Fj}&{\widetilde Z}_{FE}\cr
              {\widetilde Z}_{lF}&{\widetilde Z}_{lj}&{\widetilde Z}_{lE}\cr
              0&0&{\widetilde Z}_{EE}\cr}\right)
\left(\matrix{R^n_F\cr
              R^n_j\cr
              R^n_E\cr}\right)_R ,
\ \ \ \ \left(l,j = 1,\cdot\cdot\cdot,n-2\right),
\label{Znew}
\eea
\bigskip
\noindent
where the suffix $R (B)$ denotes renormalized (bare) quantities.

Now we proceed to the moment sum rule for $\overline{g}_2$.
Define the matrix elements of composite operators between nucleon
states with momentum $p$ and spin $s$ by
\bea
  \langle p,s | R_F^{\sigma\mu_{1}\cdots \mu_{n-1}} |p,s \rangle
       &=& -  \frac{n-1}{n} d_n ( s^{\sigma}p^{\mu_1} - s^{\mu_1}p^{\sigma})
                    p^{\mu_2} \cdots p^{\mu_{n-1}} 
                     \label{elementF}\\
  \langle p,s | R_l^{\sigma\mu_{1}\cdots \mu_{n-1}} |p,s \rangle
       &=& - f^l_n ( s^{\sigma}p^{\mu_1} - s^{\mu_1}p^{\sigma})
                    p^{\mu_2} \cdots p^{\mu_{n-1}} \\
  \langle p,s | R_E^{\sigma\mu_{1}\cdots \mu_{n-1}} |p,s \rangle
      &=& 0 \label{elementE}.
\eea
Normalization is such that for free quark target we have 
$d_n =1$ and $f_n^l= {\cal O} (g^2 )$. It is recalled that physical matrix 
elements of the EOM operators vanish~\cite{POLI}. 
Using Eqs.(\ref{elementF}) - (\ref{elementE}),
we can write down the moment sum rule for $\overline{g}_2$ as,
\be
    M_n \equiv \int_0^1 dx x^{n-1} \overline{g}_2 (x,Q^2) =
           {{n-1}\over {2n}}  d_n E_F^n(Q^2) 
    + {1\over 2}  \sum_{l=1}^{n-2}f^l_n E_l^n(Q^2) .
\label{g2sumrule}
\ee

The coefficient functions $E_F^n(Q^2)$ and $E_l^n(Q^2)$ satisfy the following 
renormalization group equation,
\be
   \left( \mu \frac{\partial}{\partial \mu}
    + \beta (g) \frac{\partial}{\partial g} 
      -\gamma_m(g) m \frac{\partial}{\partial m} \right) E_i =
        {\widetilde \gamma}_{ji}  E_j \qquad {\rm for}\ \  i,j=F, 1, \cdots, {n-2} 
      \label{RGeq}
\ee
where $\beta (g)$ and $\gamma_m(g)$ are the QCD $\beta$ function and 
the anomalous dimension of mass operator, respectively. 
The anomalous dimension matrix ${\widetilde \gamma}_{ij}$
of the composite operators $R^n_F$ and $R^n_l$ with $l=1,\cdots,n-2$  is defined as 
\be
  {\widetilde \gamma}_{ij}=\left[{\widetilde Z}^{-1} 
     \mu \frac{\partial {\widetilde Z}}{\partial \mu} \right]_{ij}
    \qquad {\rm for}\ \  i,j=F, 1, \cdots, {n-2}~.   \label{AnoDi}
\ee
Note that the anomalous dimension matrix which appears in Eq.(\ref{RGeq}) 
is a transposed one. This comes from our 
convention of defining renormalization constants 
and anomalous dimensions of the operators in Eqs.(\ref{Znew}) and 
(\ref{AnoDi}). 

In the leading-logarithmic approximation, the solutions of the RG equations 
in Eq.(\ref{RGeq}) are given as follows~\cite{Muta}:
\be
    E^n_i(Q^2)=\left[{\rm exp} \left\{ \frac{{\widetilde \gamma}^{(0)n}}{2\beta_0} 
    {\rm ln} \left(\frac{\alpha(Q^2)}{\alpha(\mu^2)}\right) \right\} \right]_{Fi}
  \qquad {\rm for}\ \  i=F, 1, \cdots, {n-2}.   \label{Solution}
\ee
where $\alpha(Q^2)$ is the QCD running coupling constant, 
$\beta_0$ and ${\widetilde \gamma}^{(0)n}$ are, respectively,  
one-loop coefficients of the $\beta$ function and anomalous dimension matrix, 
\bea
    \beta~(g)&=&-\beta_0 g^3 + {\cal O}(g^5)~, \qquad 
        \beta_0=\frac{1}{(4\pi)^2}\frac{11 N_c-2n_f}{3} \\
   {\widetilde \gamma}^n_{ij}~(g)&=&{\widetilde \gamma}^{(0)n}_{ij} g^2 +
   {\cal O}(g^4)~,   \label{gammazero}
\eea
with $n_f$ being the number of flavours, 
and we have used the fact that $E^n_F(\mu^2)=1$ and $E^n_l(\mu^2)=0$ (for 
$l=1,\cdots,n-2$) at the lowest-order. 

Now we need the information on the  anomalous dimensions 
$({\widetilde \gamma}^{(0)n})_{Fi}$ (for $i=F,1,\cdots,n-2$).  
We can get it without embarking on a new calculation of the relevant
Feynman diagrams. We utilize the existing results on 
the anomalous dimension matrix for the operators 
$R_l$,$R_m$ and $R_E$. 
In the case of the conventional choice of $R_k$,$R_m$ and $R_E$ 
as independent operators, the renormalization constant matrix 
takes a triangular form 
\bea
\left(\matrix{R^n_l\cr
              R^n_m\cr
              R^n_E\cr}\right)_B =
\left(\matrix{Z_{lj}&Z_{lm}&Z_{lE}\cr
              0&Z_{mm}&0\cr
              0&0&Z_{EE}\cr}\right)
\left(\matrix{R^n_j\cr
              R^n_m\cr
              R^n_E\cr}\right)_R ,
\ \ \ \ \left(l,j = 1,\cdot\cdot\cdot,n-2\right),
\label{Zold}
\eea
In the MS renormalization scheme $Z_{ij}$ is expressed as
\be
  Z_{ij} = \delta_{ij}- {g^2 \over 16\pi^2 \varepsilon} X_{ij}
    \ \ \ \ \left(i,j= 1,\cdot\cdot\cdot,n-2, m , E\right), 
  \label{Zij}
\ee
where $\varepsilon=(4-d)/2$ with $d$ the space-time dimension, and 
the components $X_{ij}$ have been calculated
\cite{BKL}\cite{Ratcliffe}\cite{JiChou}\cite{KTUY}. 
The following is the result on $X_{ij}$ taken from Ref.\cite{KTUY}:
\bea
 X_{lj} &=&  C_G \frac{(j+1)(j+2)}{(l+1)(l+2)(l-j)} \nonumber \\
   & &+ \, (2C_F-C_G) \left[ (-1)^{l+j}\frac{\ _{n-2}C_{j-1}}
        {\ _{n-2}C_{l-1}  }
    \frac{(n-1+l-j)}{(n-1)(l-j)}+\frac{2(-1)^{j}}{l(l+1)(l+2)}
      \ _{l}C_{j}\right]\nonumber\\
    & & \qquad \qquad \qquad \qquad \qquad \qquad \qquad \qquad \qquad
     (1\le j \le l-1), \label{eqXlj}\\
 X_{ll} &=& C_G \left(\frac{1}{l} - 
   \frac{1}{l+1}-\frac{1}{l+2}-\frac{1}{n-l}-S_{l}
    -S_{n-l-1} \right)\nonumber \\
   & &  + (2C_F - C_G)\left[\frac{1}{n-1} 
   + \frac{2(-1)^{l}}{l (l+1)(l+2)}-\frac{(-1)^{l}}{n-l}\right]
           \nonumber \\
   & & 
      + C_F \left(3-2S_{l} -2S_{n-l-1} \right),  \label{eq314}\\
 X_{lj} &=& C_G\frac{(n-1-j)(n-j)}{(n-1-l)(n-l)(j-l)} \nonumber\\ 
   & & +(2C_F - C_G) \left[ (-1)^{l+j}\frac{\ _{n-2}C_{j}}
     {\ _{n-2}C_{l}} \frac{(n-1-l+j)}{(n-1)(j-l)}
   +(-1)^{n-j}\frac{\ _{n-2-l}C_{n-2-j}}{n-l} \right] \nonumber\\
  & & \qquad \qquad \qquad \qquad \qquad \qquad \qquad \qquad \qquad
    \left( l+1 \le j \le n-2\right), \label{eq315}\\
  X_{lm} &=& \frac{4C_{F}}{n l(l+1)(l+2)} \ ,  \qquad \quad
  X_{mm}= - 4C_{F}S_{n-1} \ . \label{eqXmm}
\eea

If we impose that the renormalized and bare operators respectively 
satisfy the constraint Eq.(\ref{oprelation}), we find from Eqs.(\ref{Znew}) 
and (\ref{Zold}) that $\widetilde Z$'s are related to 
the conventional $Z$'s as follows:
\bea
   {\widetilde Z}_{FF}&=&Z_{mm}+\frac{n}{n-1} \sum_{l=1}^{n-2} (n-1-l) Z_{lm}, \\
   {\widetilde Z}_{Fj}&=& -(n-1-j){\widetilde Z}_{FF}+
             \sum_{l=1}^{n-2} (n-1-l) Z_{lj}, \\
   {\widetilde Z}_{lF}&=& \frac{n}{n-1}Z_{lm}, \\
   {\widetilde Z}_{lj}&=& Z_{lj}- \frac{n}{n-1}(n-1-j)Z_{lm},  \label{wideZil}
\eea
where  $l,j=1, \cdots, n-2$.
Using the MS scheme-rule, $1/\varepsilon \rightarrow {\rm ln}\mu^2$, 
we obtain from Eqs.(\ref{AnoDi}) and (\ref{gammazero}), 
\bea
   -8\pi^2~{\widetilde \gamma}^{(0)n}_{FF}&=&X_{mm}+\frac{n}{n-1} 
              \sum_{l=1}^{n-2} (n-1-l) X_{lm}, \\
   -8\pi^2~{\widetilde \gamma}^{(0)n}_{Fj}&=& -(n-1-j)\Biggl[
            X_{mm}+\frac{n}{n-1}\sum_{l=1}^{n-2} (n-1-l) X_{lm}\Biggr]  
    \nonumber  \\
    & & + \sum_{l=1}^{n-2} (n-1-l) X_{lj}, \\
   -8\pi^2~{\widetilde \gamma}^{(0)n}_{lF}&=& \frac{n}{n-1}X_{lm}, \\
   -8\pi^2~{\widetilde \gamma}^{(0)n}_{lj}&=& X_{lj}- 
                \frac{n}{n-1}(n-1-j)X_{lm},  \label{wideZil}\\
    &\cdots&   \ \ \ . \nonumber  
\eea
It is straightforward to calculate the above ${\widetilde \gamma}^{(0)n}_{ij}$ 
using the expressions $X_{ij}$ in Eqs.(\ref{eqXlj})-(\ref{eqXmm}).  
Especially we obtain 
\bea
 8\pi^2~{\widetilde \gamma}^{(0)n}_{FF}&=&4C_{F}\Bigl(S_{n-1}-\frac{1}{4}
                          +\frac{1}{2n}\Bigr), \\
  8\pi^2~{\widetilde \gamma}^{(0)n}_{Fj}&=& 
      -(2C_{F}-C_G)\Biggl[(n-1-j)\Biggl\{ 2S_{n-1}-S_j-S_{n-j-1}+1+\frac{1}{n}\Biggr\}
     \nonumber \\
 & &  +\sum_{l=1}^{j-1}(n-1-l)\Biggl\{ (-1)^{l+j}\frac{\ _{n-2}C_{j}}
     {\ _{n-2}C_{l}} \frac{(n-1-l+j)}{(n-1)(j-l)}
   +(-1)^{n-j}\frac{\ _{n-2-l}C_{n-2-j}}{n-l} \Biggr\}  \nonumber \\
 & & +(n-1-j) \Biggl\{ \frac{1}{n-1} 
   + \frac{2(-1)^{j}}{j (j+1)(j+2)}-\frac{(-1)^{j}}{n-j}  \Biggr\} \nonumber \\
& & +\sum_{l=j+1}^{n-2}(n-1-l)\Biggl\{(-1)^{l+j}\frac{\ _{n-2}C_{j-1}}
        {\ _{n-2}C_{l-1}  }
    \frac{(n-1+l-j)}{(n-1)(l-j)}+\frac{2(-1)^{j}}{l(l+1)(l+2)}
      \ _{l}C_{j}  \Biggr\} 
\Biggr] \nonumber  \\
& & \qquad \qquad \qquad \qquad {\rm for}\ \ j=1, \dots, n-2
\eea

Now we see that the mixing anomalous dimension ${\widetilde \gamma}^{(0)n}_{Fj}$ 
turns out to be proportinal to $ (2C_{F}-C_G)$. Since
\be
     C_{F}=\frac{N_C^2-1}{2N_C}, \qquad \quad  C_G=N_C   ,
\ee
we have $2C_{F}=C_G$ and thus ${\widetilde \gamma}^{(0)n}_{Fj}=0$ in the 
$N_C \rightarrow \infty$ limit. Then Eq.(\ref{Solution}) gives,
\bea
       E^n_F (Q^2)&=&\Biggl[\frac{\alpha(Q^2)}{\alpha(\mu^2)} 
    \Biggr]^{{\widetilde \gamma}^{(0)n}_{FF}/2\beta_0} ~,   \\ 
      E^n_l (Q^2)&=&0   \qquad \qquad {\rm for} \quad l=1, \cdots, n-2~.  
\eea
Returning to Eq.(\ref{g2sumrule}), we find that, at $N_C$ going to infinity, 
the moment sum rule for $\overline{g}_2$ takes a simple form as follows:
\be
   \int_0^1 dx x^{n-1} \overline{g}_2 (x,Q^2) ={{n-1}\over {2n}}  d_n 
   \Biggl[\frac{\alpha(Q^2)}{\alpha(\mu^2)} 
    \Biggr]^{{\widetilde \gamma}^{(0)n}_{FF}/2\beta_0}~.  \label{result}
\ee
with
\be
     \frac{{\widetilde \gamma}^{(0)n}_{FF}}{2\beta_0}= 
       \frac{2 N_C \Bigl[ S_{n-1} -\frac{1}{4}+\frac{1}{2n} \Bigr]}
        {\frac{1}{3}\Bigl[11 N_C - 2 n_f   \Bigr]} ~.
\ee
In other words, at large $N_C$, the operators $R_l^{\sigma\mu_{1}\cdots \mu_{n-1}}$ 
involving the gluon fields 
decouple from the evolution equation of $\overline{g}_2$  and 
the whole contribution is represented by one type of  
operators $ R_F^{\sigma\mu_{1}\cdots \mu_{n-1}}$. 
With the substitution $C_F=N_C/2$ and $n=j+1$, the anomalous dimension 
$8\pi^2 {\widetilde \gamma}^{(0)n}_{FF}$ coincides with Eq.(18) of 
Ref.\cite{ABH}. This completes the reproduction, 
in the framework of OPE and RG, of the ABH result on $\overline{g}_2$.

It should be emphasized that we have reproduced the ABH result
without assuming massless quarks. A question expected to come up immediately 
is that the replacement of the mass-dependent operator $R_m^n$ with $R_F^n$ may 
be equivalent to working with massless quarks. The answer is no. Indeed it can be  
shown that even when we include the mass-dependent operator $R_m^n$ among the 
independent operator bases we reach the same conclusion. 
Let us take, for an example, $R_F^n$, $R_l^n$ (with $l=2,\cdots,n-2$), $R_m^n$ and 
$R_E^n$ as independent operators replacing one quark-gluon operator $R_{l=1}^n$ 
with $R_F^n$. With this choice of new operator bases, the moment sum rule for 
$\overline{g}_2$ is written in terms of the coefficient functions  
${\widehat E}_F^n(Q^2)$, ${\widehat E}_l^n(Q^2)$ with $l=2,\cdots,n-2$, and 
${\widehat E}_m^n(Q^2)$. 
The renormalization constants for these operators are written as 
\bigskip
\bea
\left(\matrix{R^n_F\cr
              R^n_l\cr
              R^n_m\cr
              R^n_E\cr}\right)_B =
\left(\matrix{{\widehat Z}_{FF}&{\widehat Z}_{Fj}&{\widehat Z}_{Fm}&{\widehat Z}_{FE}\cr
          {\widehat Z}_{lF}&{\widehat Z}_{lj}&{\widehat Z}_{lm}&{\widehat Z}_{lE}\cr
    0&0&{\widehat Z}_{mm}&0\cr
              0&0&0&{\widehat Z}_{EE}\cr}\right)
\left(\matrix{R^n_F\cr
              R^n_j\cr
              R^n_m\cr
              R^n_E\cr}\right)_R ,
\ \ \ \ \left(l,j = 2,\cdot\cdot\cdot,n-2\right).
\label{Znewnew}
\eea

\bigskip
\noindent
Again imposing that the renormalized and bare operators respectively 
satisfy the constraint Eq.(\ref{oprelation}), we find 
that $\widehat Z$'s are related to conventional $Z$'s as follows:
\bea
   {\widehat Z}_{FF}&=&\frac{1}{n-2}\sum_{l=1}^{n-2} (n-1-l) Z_{l1} \\
   {\widehat Z}_{Fj}&=& -(n-1-j){\widehat Z}_{FF}+
             \sum_{l=1}^{n-2} (n-1-l) Z_{lj},  
    \ \ \ \ \left(j = 2,\cdot\cdot\cdot,n-2\right)  \\
  {\widehat Z}_{Fm}&=&-\frac{n-1}{n}{\widehat Z}_{FF}+\frac{n-1}{n}Z_{mm} +
 \sum_{l=1}^{n-2} (n-1-l) Z_{lm}
\eea
Then it is easy to obtain the following one-loop coefficients of 
the relevant anomalous dimensions 
\bea
 8\pi^2~{\widehat \gamma}^{(0)n}_{FF}&=&4C_{F}\Bigl(S_{n-1}-\frac{1}{4}
                         +\frac{1}{2n}\Bigr)   \nonumber   \\
 & & \qquad \qquad +~ {\rm terms\ proportional\ to\ }(2C_{F}-C_G), \\
   8\pi^2~{\widehat \gamma}^{(0)n}_{Fj}&\propto& 
      (2C_{F}-C_G) \qquad \quad {\rm for}\ \ j=2, \dots, n-2  \\
   8\pi^2~{\widehat \gamma}^{(0)n}_{Fm}&\propto& 
      (2C_{F}-C_G)
\eea
Inserting these anomalous dimensions to the  
solutions of the RG equations for the coefficient functions 
${\widehat E}_F^n(Q^2)$, ${\widehat E}_l^n(Q^2)$ ($l=2,\cdots,n-2$) and 
${\widehat E}_m^n(Q^2)$~,
\be
    {\widehat E}^n_i(Q^2)=\left[{\rm exp} \left\{ \frac{{\widehat \gamma}^{(0)n}}{2\beta_0} 
    {\rm ln} \left(\frac{\alpha(Q^2)}{\alpha(\mu^2)}\right) \right\} \right]_{Fi}
  \qquad {\rm for}\ \  i=F, 2, \cdots, {n-2}, m~,   \label{NewSolution}
\ee
we obtain in the large $N_C$ limit
\bea
       {\widehat E}^n_F (Q^2)&=&\Biggl[\frac{\alpha(Q^2)}{\alpha(\mu^2)} 
    \Biggr]^{{\widetilde \gamma}^{(0)n}_{FF}/2\beta_0} = E^n_F (Q^2)~,   \\ 
      {\widehat E}^n_l (Q^2)&=&0   \qquad \qquad {\rm for} \quad l=2, \cdots, n-2 \\
     {\widehat E}^n_m (Q^2)&=&0 .  
\eea
Thus we reach the same conclusion Eq.(\ref{result}) even when we include 
the mass-dependent operators among the independent operator bases.

A few comments are in order. 
Firstly, the twist-3 quark-gluon operators $R_l^n$ decouple 
from the evolution equation for $\overline{g}_2$ at large $N_C$.
This might be explained by an argument on quark condensate~\cite{ShuVain}.
A hint is that the mixing anomalous dimensions 
${\widetilde \gamma}^{(0)n}_{Fj}$ turn out to be proportinal to $(2C_{F}-C_G)$.
There are two types in the products of  colour matrices 
entering into the calculation of 
anomalous dimensions for the flavour nonsinglet $\overline{g}_2$:
\bea
    T^b T^a T^b &=& (C_F-\frac{1}{2}C_G) T^a = -\frac{1}{2N_C} ~T^a \\
    T^b T^b T^a &=& C_F T^a =\frac{1}{2}\Bigl(N_C-\frac{1}{N_C}
                         \Bigr)  ~T^a  ~.
\eea
It is argued in Ref.\cite{ShuVain} that the quark condensate contains all 
colours and at large $N_C$ the condensate polarization becomes small 
and that the combination $T^b T^a T^b$ is connected with condensate 
polarization effects.

Secondly, we have chozen  particular sets of the independent 
operators and reached a simple form for the moments of $\overline{g}_2$ 
in the large $N_C$ limit. 
However, arbitrariness in the choice of the operator bases should not 
enter into physical quantities~\cite{KUY}. A different choice of the 
operator bases leads to different forms for the anomalous dimension matrix and 
the coefficient functions. Recall that the constraint, Eq.(\ref{oprelation}),  
gives a relation among 
the tree-level coefficient functions and also a relation among the matrix 
elements of the operators. After diagonalizing the anomalous dimension matrix 
and using these relations, we can arrive at the same conclusion for 
the moments of $\overline{g}_2$ in the $N_C \rightarrow \infty$ limit. 
What we did in this paper is that we choze particular sets of 
bases from the beginning which include an operator that represents 
the whole contribution to $\overline{g}_2$ for large $N_C$. 

Finally, the nucleon has other twist-3 distributions, namely, chiral-odd 
distributions $h_L(x,Q^2)$ and $e(x,Q^2)$~\cite{JJ}. Just like 
the $\overline{g}_2$ case, the $Q^2$ evolutions of flavour 
nonsinglet $h_L(x,Q^2)$ and $e(x,Q^2)$ turn out to be quite complicated due to 
mixing with quark-gluon operators, the number of which increases with spin. 
However, it has been proved~\cite{BBKT} that  in the large $N_C$ limit 
these twist-3 distributions also obey a simple DGLAP equation.
The proof holds true only  when we work with massless quarks.

\bigskip
\bigskip
\bigskip
\bigskip

This work was inspired by an interesting talk given by Y. Koike 
at {\it International Symposium on QCD Corrections and New Physics}, 
27-29 October 1997, Hiroshima.  
I would like to thank him and also thank the organizer of the Symposium, 
J. Kodaira.  The discussions with Y. Koike and T. Uematsu on the 
twist-3 operators in the large $N_C$ limit were indispensable for the 
completion of this paper and are happily acknowledged. 
This work is supported in part by the Monbusho Grant-in-Aid for 
Scientific Research No. (C)(2)-09640342.



\newpage
\begin{thebibliography}{99}


\bibitem{review}
     See, for example, R.~L.~Jaffe, MIT-CTP-2506 and HUTP-96/A003 hep-ph/9602236.

\bibitem{ARS}
     M.~A.~Ahmed and G.~G.~Ross, {\sl Phys.Lett.} {\bf B56} (1975) 385; 
{\sl Nucl.Phys.} {\bf B111} (1976) 441;\\
     K.~Sasaki, {\sl Prog.Theor.Phys.} {\bf 54} (1975) 1816.

\bibitem{KOD1}
     J.~Kodaira, S.~Matsuda, K.~Sasaki and T.~Uematsu, {\sl Nucl.Phys.}
                                {\bf B159} (1979) 99.
\bibitem{KOD2}
     J.~Kodaira, S.~Matsuda, T.~Muta, K.~Sasaki and T.~Uematsu,
                    {\sl Phys.Rev.} {\bf D20} (1979) 627.

\bibitem{g1}
    J.~Ashman {\it et al.}, {\sl Nucl.Phys.} {\bf B328} (1989)1; 
    D.~Adams {\it et al.}, {\sl Phys.Lett.} {\bf B329} (1994) 399;
         {\bf B357} (1995) 248; {\bf B396} (1997) 338; 
    K.~Abe {\it et al.}, {\sl Phys.Rev.Lett.} {\bf 74} (1995) 346;
        {\bf 75} (1995) 25; {\bf 79} (1997) 26;  
         {\sl Phys.Lett.} {\bf B364} (1995) 61; {\bf B404} (1997) 377; 
                     {\bf B405} (1997) 180; 
   K.~Ackerstaff {\it et al.}, {\sl Phys.Lett.} {\bf B404} (1997) 383.


\bibitem{g2}
      P.~L.~Anthony {\it et al.}, {\sl Phys.Rev.Lett.} {\bf 71} (1993) 959;
                  {\sl Phys.Rev.} {\bf D54} (1996) 6620;  
      D.~Adams {\it et al.},{\sl Phys.Lett.} {\bf B336} (1994) 125; 
      K.~Abe {\it et al.}, {\sl Phys.Rev.Lett.} {\bf 76} (1996) 587; 
         {\sl Phys.Lett.} {\bf B404} (1997) 377.


\bibitem{ShuVain}
     E.~V.~Shuryak and A.~I.~Vainshtein, {\sl Nucl.Phys.}
                    {\bf B199} (1982) 951; {\bf B201} (1982) 141.

\bibitem{BKL}
      A.~P.~Bukhvostov, E.~A.~Kuraev and L.~N.~Lipatov, 
    {\sl Sov. J. Nucl. Phys.}{\bf 38} (1983) 263; 39 (1984) 121;
        {\sl JETP Letters} {\bf 37} (1984) 482; 
        {\sl Sov. Phys. JETP} {\bf 60} (1984) 22.

\bibitem{Ratcliffe}
     P.~G.~Ratcliffe, {\sl Nucl. Phys.} {\bf B264} (1986) 493.

\bibitem{BB}
     I.~I.~Balitsky and V.M.Braun, {\sl Nucl. Phys.} {\bf B311} (1989) 541.

\bibitem{JiChou}
     X.~Ji and C.~Chou, {\sl Phys.Rev.} {\bf D42} (1990) 3637. 

\bibitem{Jaffe}
     R.~L.~Jaffe, {\sl Comments Nucl.Part.Phys.} {\bf 19} (1990) 239.

\bibitem{JaffeJi}
    R.~L.~Jaffe and X.~Ji, {\sl Phys.Rev.} {\bf D43} (1991) 724.

\bibitem{KMSU}
     J.~Kodaira, S.~Matsuda,T.~Uematsu and  K.~Sasaki, 
       {\sl Phys.Lett.} {\bf B345} (1995) 527.

\bibitem{ABH} 
     A.~Ali, V.~M.~Braun and G.~Hiller, {\sl Phys.Lett.}{\bf B266}
                               (1991) 117.

\bibitem{GLAP}
      V.~N.~Gribov and L.~N.~Lipatov, {\sl Sov. J. Nucl. Phys}
                    {\bf 15} (1972) 675; 
       L.~N.~Lipatov, {\it ibid.} {\bf 20} (1975) 94; 
    Y.~L.~Dokshitzer, {\sl Sov. Phys. JETP} {\bf 46} (1977) 461; 
    G.~Altarelli and G.~Parisi, {\sl Nucl. Phys.} {\bf B126} (1977) 298.


\bibitem{KUY}
     J.~Kodaira, T.~Uematsu and Y.~Yasui,{\sl Phys.Lett.} {\bf B344} (1995) 348.


\bibitem{KTUY}
     J.~Kodaira, K.~Tanaka, T.~Uematsu and Y.~Yasui, 
           {\sl Phys.Lett.} {\bf B387} (1996) 855.

\bibitem{WW}
     W.~Wandzura and F.~Wilczek, {\sl Phys.Lett.} {\bf B172} (1977) 195.

\bibitem{POLI}
     H.~D.~Politzer, {\sl Nucl.Phys.} {\bf B172} (1980) 349.

\bibitem{Muta}
     T.~Muta, {\it Foundations of Quantum Chromodynamics} 
    (World Scientific, Singapore, 1987).

\bibitem{JJ}
     R.~L.~Jaffe and X.~Ji, {\sl Phys.Rev.Lett.} {\bf 67} (1991) 552; 
      {\sl Nucl.Phys.} {\bf B375} (1992) 527.

\bibitem{BBKT}
     I.~I.~Balitsky, V.~M.~Braun, Y.~Koike and K.~Tanaka, 
      {\sl Phys.Rev.Lett.} {\bf 77} (1996) 3078; 
     Y.~Koike and K.~Tanaka, {\sl Phys.Rev.} {\bf D51} (1995) 6125;
    Y.~Koike and N.~Nishiyama,  {\sl Phys.Rev.} {\bf D55} (1997) 3068.








\end{thebibliography}
\end{document}